\documentclass[twocolumn,showpacs,aps,prd,superscriptaddress,paperletter]
{revtex4}

\usepackage{graphicx}
\usepackage{dcolumn}
\usepackage{amsmath}
\usepackage{epsfig}
\usepackage{multirow}

\input pubboard/babarsym

\def\PM     {\ensuremath{\pm}\xspace}

\def\mX     {\ensuremath{m_X}\xspace}

\def\PDG00{{\it Particle Data Group 2000}}
\def\ksdk   {\ensuremath{\KS \to \pipi}\xspace}
\def\pizdk  {\ensuremath{\piz \to \gaga}\xspace}
\def\phidk  {\ensuremath{\phi \to \Kp\Km}\xspace}

\def\KKbar  {\ensuremath{\kaon \Kb}\xspace}

\def\etacK  {\ensuremath{\B \to \etac \kaon}\xspace}
\def\etacKN {\ensuremath{\Bz \to \etac \KS}\xspace}
\def\etacKn {\ensuremath{\Bz \to \etac \Kz}\xspace}

\def\etacKP {\ensuremath{\Bp \to \etac \Kp}\xspace}

\def\jpsiK  {\ensuremath{\B \to \jpsi \kaon}\xspace}

\def\eTwoPhi {\ensuremath{\etac \to \phi\phi}\xspace}
\def\eKKPi   {\ensuremath{\etac \to \KpKm \piz}\xspace}
\def\eKsKPi  {\ensuremath{\etac \to \KS \Kpm \pimp}\xspace}
\def\eKKbarPi {\ensuremath{\etac \to \KKbar \pi}\xspace}
\def\eKzKPi  {\ensuremath{\etac \to \Kz \Km \pip}\xspace}

\def\TwoPhi {\ensuremath{\phi\phi}\xspace}
\def\KKPi   {\ensuremath{\KpKm \piz}\xspace}
\def\KsKPi  {\ensuremath{\KS \Kpm \pimp}\xspace}

\def\shorteFourK  {\ensuremath{\etac \to 2 (\KpKm)}\xspace}
\def\shortFourK  {\ensuremath{ 2( \KpKm ) }\xspace}

\newcommand{\BABARPubYear}    {03}
\newcommand{\BABARPubNumber}  {043}

\newcommand{\SLACPubNumber} {10368}

\def\figurebox#1#2#3{%
    \def\arg{#3}%
    \ifx\arg\empty
    {\hfill\vbox{\hsize#2\hrule\hbox to #2{\vrule\hfill\vbox to 
#1{\hsize#2\vfill}\vrule}\hrule}\hfill}%
    \else
    {\hfill\epsfbox{#3}\hfill}%
    \fi}

\begin{document}

\preprint{\babar-PUB-\BABARPubYear/\BABARPubNumber} 
\preprint{SLAC-PUB-\SLACPubNumber} 

\begin{flushleft}
\babar-PUB-\BABARPubYear/\BABARPubNumber\\
SLAC-PUB-\SLACPubNumber\\
\end{flushleft}

\title{
{\large \bf
Branching Fraction Measurements of \boldmath{\etacK} Decays} 
}

%
\author{B.~Aubert}
\author{R.~Barate}
\author{D.~Boutigny}
\author{F.~Couderc}
\author{J.-M.~Gaillard}
\author{A.~Hicheur}
\author{Y.~Karyotakis}
\author{J.~P.~Lees}
\author{V.~Tisserand}
\author{A.~Zghiche}
\affiliation{Laboratoire de Physique des Particules, F-74941 Annecy-le-Vieux, France }
% ...
\collaboration{The \babar\ Collaboration}
\noaffiliation

\date{\today}%

\begin{abstract}

We study the decays \etacKP and \etacKn, where the \etac\ is 
reconstructed in the \KsKPi and \KKPi decay modes. Results are based on a 
sample of 86 million \BB pairs collected with the \babar\ detector at the 
SLAC \epem \BF. We measure the branching fractions
$\BR(\etacKP) = (1.34 \pm 0.09 \pm 0.13 \pm 0.41) \times 10^{-3} $ and
$\BR(\etacKn) = (1.18 \pm 0.16 \pm 0.13 \pm 0.37) \times 10^{-3} $,
where the first error is statistical, the second is systematic, and the 
third reflects the \etac\ branching fraction uncertainty. In addition, we 
search for \etacK events with \shorteFourK and \eTwoPhi and determine the 
\etac\ decay branching fraction ratios 
$\BR(\shorteFourK)/\BR(\eKKbarPi) = (2.3 \pm 0.7 \pm 0.6) \times 10^{-2}$ 
and
$\BR(\eTwoPhi)/\BR(\eKKbarPi) = (5.5 \pm 1.4 \pm 0.5) \times 10^{-2}$.

\end{abstract}


\maketitle

The decay \etacK is used to measure \stwob\cite{sin2bBabar,sin2bBelle}, but is
interesting dynamically as well. The ratio of its decay rate to that of \jpsiK
reflects the underlying strong dynamics and can be used to check models
of heavy quark systems~\cite{QCD,RK1,RK2,RK3,RK4}. The strong decay should
be isospin invariant, an expectation that can be checked and then used
to combine results for higher precision. It is therefore 
interesting to measure accurately the branching fractions for \etacKn and 
\etacKP~\cite{Conjunote}.

We use data collected with the \babar\ detector at the \pep2 
energy-asymmetric \epem storage rings. The data sample contains 
$86.1 \times 10^6$ \BB pairs, corresponding to an integrated luminosity of 
79.4 \invfb taken at a center-of-mass energy equivalent to the mass of the 
\FourS resonance. An additional 9.6 \invfb\ of data, collected 40 MeV below 
the resonance, is used to study the background from light quark and \ccbar 
production.

A detailed description of the \babar\ detector can be found 
elsewhere~\cite{NIM}; only detector components relevant to this analysis 
are mentioned here. Charged-particle trajectories are measured by a 
five-layer double-sided silicon vertex tracker (SVT) and a 40-layer drift 
chamber (DCH), operating in the field of a 1.5-T solenoid. Charged 
particles are identified by combining measurements of ionization energy 
loss (\dedx) in the DCH and SVT with angular information from a detector of 
internally reflected Cherenkov light (DIRC). Photons are identified as 
isolated electromagnetic showers in a CsI(Tl) electromagnetic calorimeter. 

In this analysis, \etac\ mesons are reconstructed in the \KsKPi, \KKPi, 
\shortFourK and \TwoPhi decay modes. Candidates for \KS are identified 
through the decay \ksdk, $\phi$ candidates through \phidk and \piz 
candidates through \pizdk. Note that \etac decays to \shortFourK include 
both non-resonant and resonant (\TwoPhi, $\phi\Kp\Km$) components, so we 
expect a partial overlap of the \shortFourK and \TwoPhi samples. 

We require that charged tracks, other than those used to reconstruct \ksdk 
candidates, have a minimum transverse momentum of 0.1~\gevc, and that they 
originate from the interaction point to within 10~cm along the beam direction 
and 1.5~cm in the transverse plane. The ``fast'' kaon candidate at the 
two-body \Bp decay vertex is required to have at least 12 hits in the drift 
chamber and to have a momentum in the \FourS rest frame larger than 
1.5~\gevc. 

The cuts used to select $K^+$, \KS, $\phi$ and \piz candidates 
from \etac\ decays are described below. They are 
optimized to maximize the statistical sensitivity of the signal, defined as 
$S/\sqrt{S+B}$, with $S$ and $B$ being the estimated numbers of signal and 
combinatorial-background events.

All charged-kaon candidates are required to have momentum greater than 
250 \mevc and a polar angle between 0.35 and 2.54 rad with respect to the 
detector axis, to restrict them to a fiducial region where the particle 
identification performance can be determined with small uncertainty. Kaon 
identification is based on a neural network (NN) algorithm that combines 
information from the DCH, the SVT, and the DIRC. Particle identification 
criteria are crucial for background suppression, especially for 
\shorteFourK decays. In this channel, three of the four kaons must pass a 
tight cut on the NN output variable. Less restrictive requirements on the 
NN signature are used for identifying the fourth kaon from \shorteFourK 
candidates, the charged kaons in the other \etac\ decay modes, and the fast 
kaon from $B^+$ decays. The kaon-identification efficiency depends on the 
momentum and polar angle of the track, as well as on the chosen NN cut. For 
the tightest selection above, the average kaon efficiency exceeds 85\%; the 
corresponding pion-rejection efficiency is about 98\%.

We reconstruct \KS candidates from pairs of oppositely charged tracks 
fitted to a common vertex. We require the \KS candidate from the \B (\etac) 
decay to have a reconstructed mass within 13 (16)~\mevcc of the 
\KS mass~\cite{PDG}. Furthermore, the cosine of the opening 
angle between the flight direction and the momentum vector of the \KS 
candidate is required to be larger than 0.9995 (0.9930), and the flight 
distance from the \B vertex larger than four times its error.

We reconstruct $\phi$ candidates from pairs of oppositely charged kaons 
with an invariant mass within 14 \mevcc of the $\phi$ 
mass~\cite{PDG}.

We use pairs of photons to reconstruct \pizdk candidates, requiring a 
minimum energy of 120~MeV for one photon and 80~MeV for the other. The 
reconstructed \gaga mass is required to lie within 18~\mevcc of the
\piz mass~\cite{PDG}.

We reconstruct \etac\ candidates by fitting the appropriate combination of 
charged tracks, \KS, $\phi$, or \piz candidates to a common vertex. 
Neutral or charged \B candidates are formed from 
reconstructed \etac and \KS or \Kp candidates.
In reconstructing the \B decay chain, the measured momentum vector of each 
intermediate particle is determined by refitting the momenta of its daughters, 
constraining the mass to the nominal mass of the particle and 
requiring that the decay products originate from a common point. 
In the case of the \etac, only the geometrical vertex constraint is 
applied because of the large intrinsic width of the resonance. Charmonium 
candidates are accepted if they have an invariant mass between
2.7 and 3.3 \gevcc. Note that this procedure also reconstructs 
\jpsi\ decays, which are used to measure the mass resolution and for other 
cross-checks. 

We use a Fisher discriminant to suppress \epem\to\qqbar 
background processes. Our Fisher discriminant is a linear 
combination of 18 variables, the most important of which are the 
normalized second Fox-Wolfram moment and the angle between the thrust axis 
of the \B candidate and that of the rest of the event. Also contributing 
are the energy flow in nine 10$^{\mathrm o}$ polar angle intervals 
coaxial around the \etac 
direction in the center-of-mass frame~\cite{FisherCLEO}, the polar angles 
of the \B candidate and of the overall thrust axis, and other event-shape
variables that distinguish between \BB events and continuum background. 
The discriminant is tuned on simulated signal events 
and on off-resonance data to achieve maximum separation between signal 
and continuum background. 
Fisher coefficients are determined individually 
for each \etac\ decay mode, and threshold values are set as part of the 
cut-optimization procedure described earlier.

We select \B candidates using two nearly independent kinematic variables:
\mes, the beam-energy--substituted mass,
and \DeltaE, the difference between the energy of the \B and 
the beam energy in the center-of-mass frame~\cite{NIM}.
The \mes resolution 
is 2.6~\mevcc, dominated by the beam-energy spread. The 
\DeltaE\ resolution varies from 15 to 28~\mev, depending on the \etac decay 
mode. Signal events are expected to have \mes close to the \B 
mass and \DeltaE close to zero. Our selection requires 
$5.2<\mes<5.3$~\gevcc. After all the cuts listed above, 10--25\% of the 
selected events, depending on the \etac\ decay channel, contain more than 
one \etacK candidate in a \DeltaE window $\pm 250$~\mev wide; we then 
retain only the candidate with the smallest value of $|\DeltaE|$.
We have verified with simulated events that 
this procedure selects the correct candidate in 90--98\% of the cases, 
and that it does not bias the measurement.
Finally, we require candidates 
to lie within an optimized interval of \DeltaE that varies from $\pm$35 to 
$\pm$70\mev, depending on the decay mode. 

Events surviving the full selection chain originate from four 
different sources: \etacK decays, i.e., the signal; \jpsiK decays,
with the \jpsi decaying into the same final state as the \etac;
a combinatorial background, arising from random track combinations in continuum 
and in \BB final states; and a background component from other \B decays to the same
final state particles as the \etacK decay mode under consideration.
The last background component can contribute
events that cluster at the signal peak in \mes and \DeltaE and is therefore
termed ``peaking background''. 

Examples of peaking background for \etacK(\eKsKPi) 
are $B^+\rightarrow K^{*-} K^+ K^+$ ($K^{*-}\rightarrow\KS\pi^-$) or 
$\Bz\rightarrow K^{*0} \KS \KS$ ($K^{*0}\rightarrow K^+\pi^-$). In the 
particular case of the decay \etacKP(\eKsKPi), another important source of
peaking background comes from $\Bp\rightarrow\Dzb\KS\Kp$  
($\Dzb\rightarrow\Kp\pim$). For this $B$ decay mode 
therefore, candidates with a $\Kp\pim$ invariant mass within 15~MeV 
($3\sigma$) of the $\Dz$ mass are explicitly vetoed. Other processes, 
such as nonresonant \B decays to the selected final state, whose branching 
fractions are not well-known, can also contribute.
The mass \mX of the system recoiling against the 
fast kaon is used to separate \etacK and \jpsiK events, which peak at the 
mass of the corresponding charmonium system, from the peaking (in \mes and \DeltaE) background, 
which is expected to exhibit a linear dependence on \mX. 
This assumption is verified with large samples
of simulated \BB events. 
These studies also show that 
inclusive \B decays into \etac\ and potential cross-feed among different 
\etac\ decay modes are negligible after the event selection. 

The number of signal events is determined from an unbinned 
maximum-likelihood fit to the joint \mes and \mX distribution. Four 
hypotheses are considered to build the 2-D likelihood function: \etacK 
signal, modeled by the product of a Gaussian resolution function in \mes 
and of a non-relativistic Breit-Wigner function convoluted with a Gaussian 
resolution function in \mX; \jpsiK component, given by the product of 
Gaussian resolution functions in \mes and \mX; combinatorial background, 
modeled by an ``ARGUS'' endpoint function in \mes~\cite{ARGUS}, and a linear 
function in \mX; and peaking background, described by a function linear in 
\mX and Gaussian in \mes. 
The widths of the \mX and \mes resolution 
functions, and the mean value of the \mes distribution are free parameters 
common to the \etac and \jpsi probability density functions (p.d.f.). The 
latter two parameters also determine the \mes dependence of the 
peaking-background p.d.f., reflecting the evidence that this background is 
dominated by \B decays to the same final states as the signal. We set 
the \etac and \jpsi masses to their world-average 
values~\cite{PDG}, the endpoint of the combinatorial background function
to 5.29 \gevcc, and the \etac\ width to the value recently measured by 
\babar~\cite{gamma}. All other parameters and the number of events in the 
different components are determined by the fit, which is performed 
separately for each decay channel. 

For \etacK modes with \eKKbarPi, candidates are weighted to take into 
account small efficiency variations across the \etac\ Dalitz plot. The 
weighting procedure compensates for any resonant structure in \etac\ 
three-body decays unaccounted for by the simulated phase-space distribution, 
which is uniform over the Dalitz plot. Since all weights are close to one, 
they do not affect the shape of the different components and have only a 
marginal influence (0.6--4\%) on the fitted event yield. Samples of 
simulated events are used to verify that the likelihood fit is unbiased.

\begin{figure}[!h]
\begin{center}
\mbox{\epsfig{file=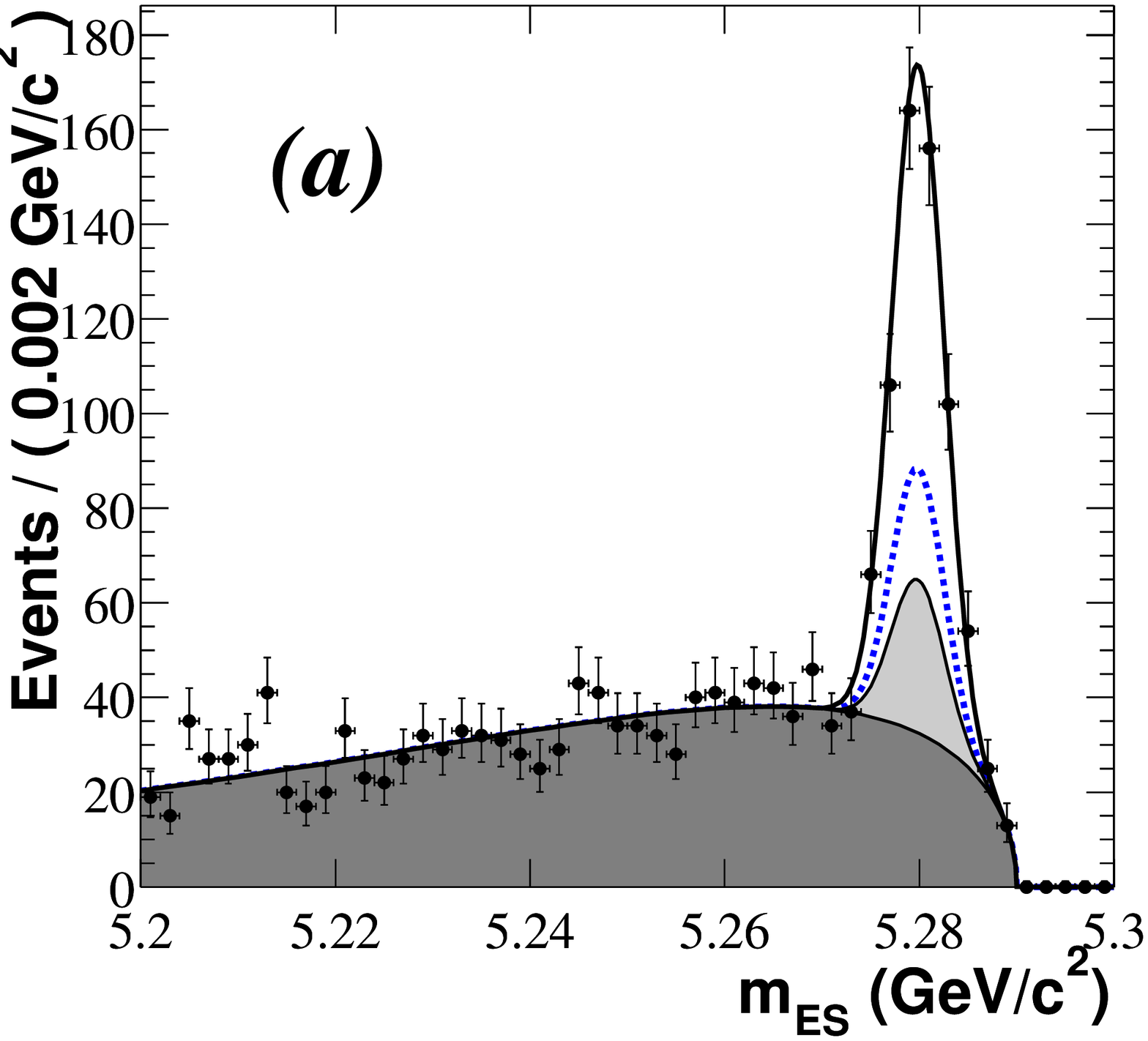,width=4.2cm}
\epsfig{file=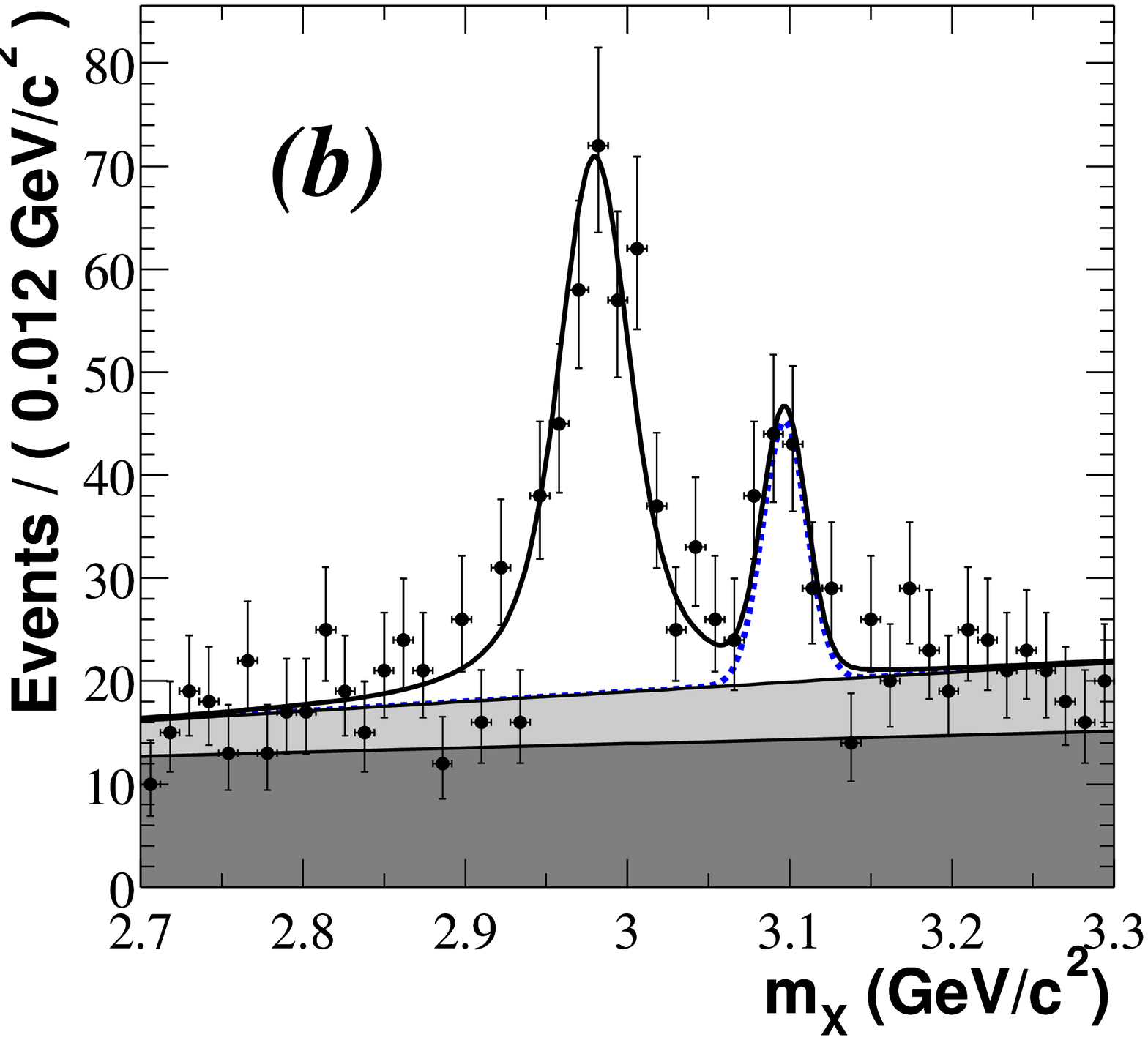,width=4.2cm}}
\mbox{\epsfig{file=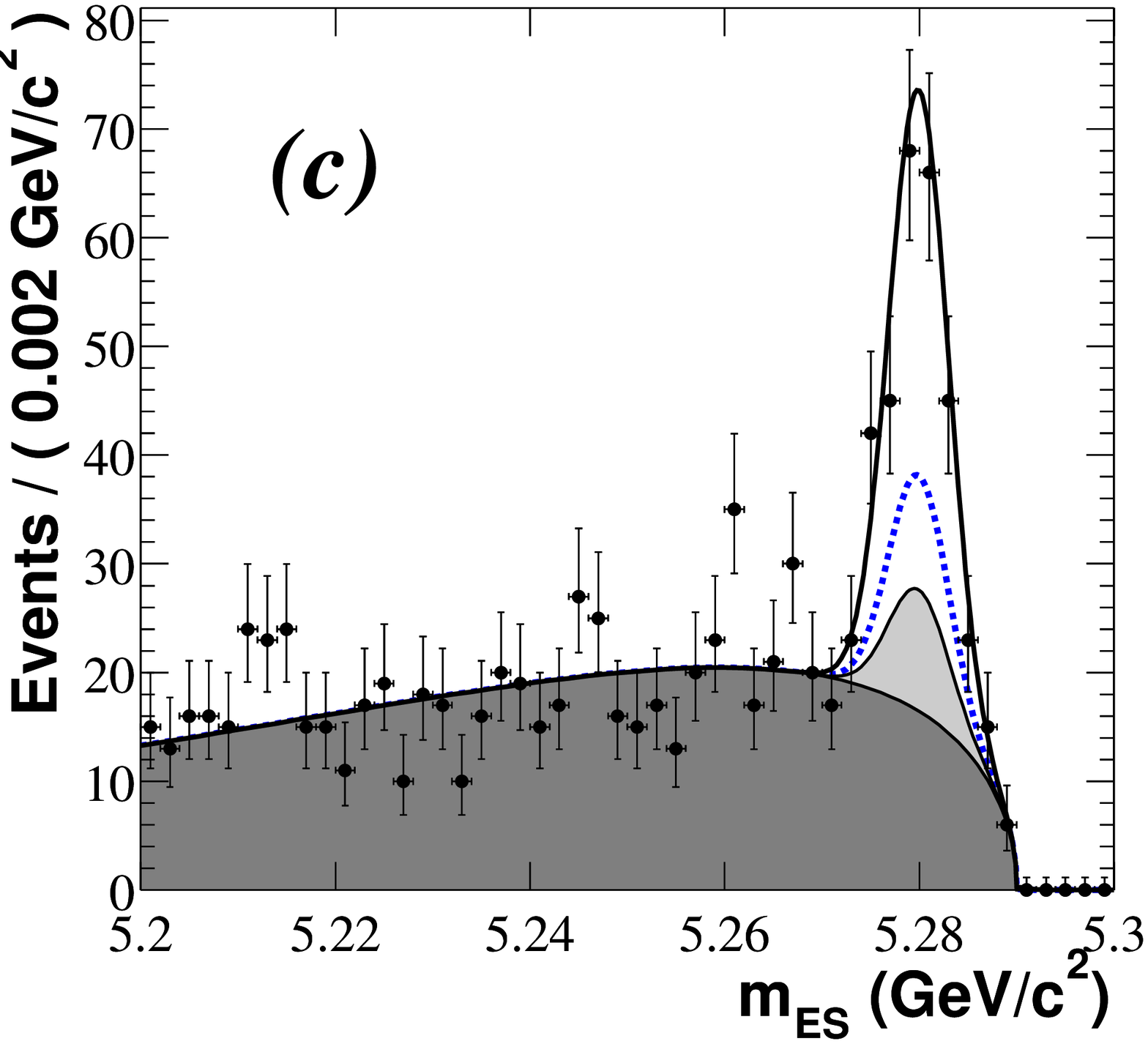,width=4.2cm}
\epsfig{file=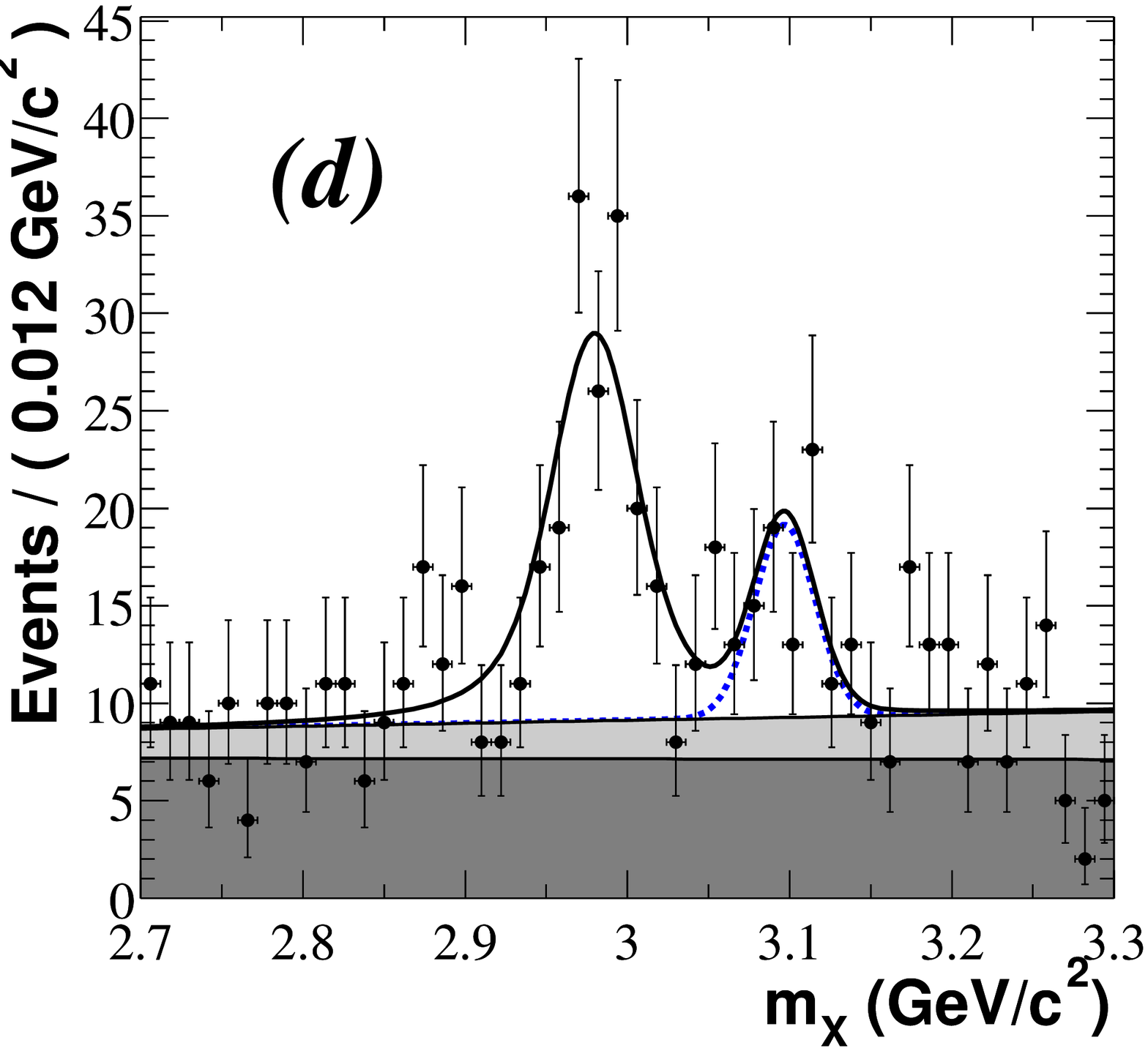,width=4.2cm}}
\mbox{\epsfig{file=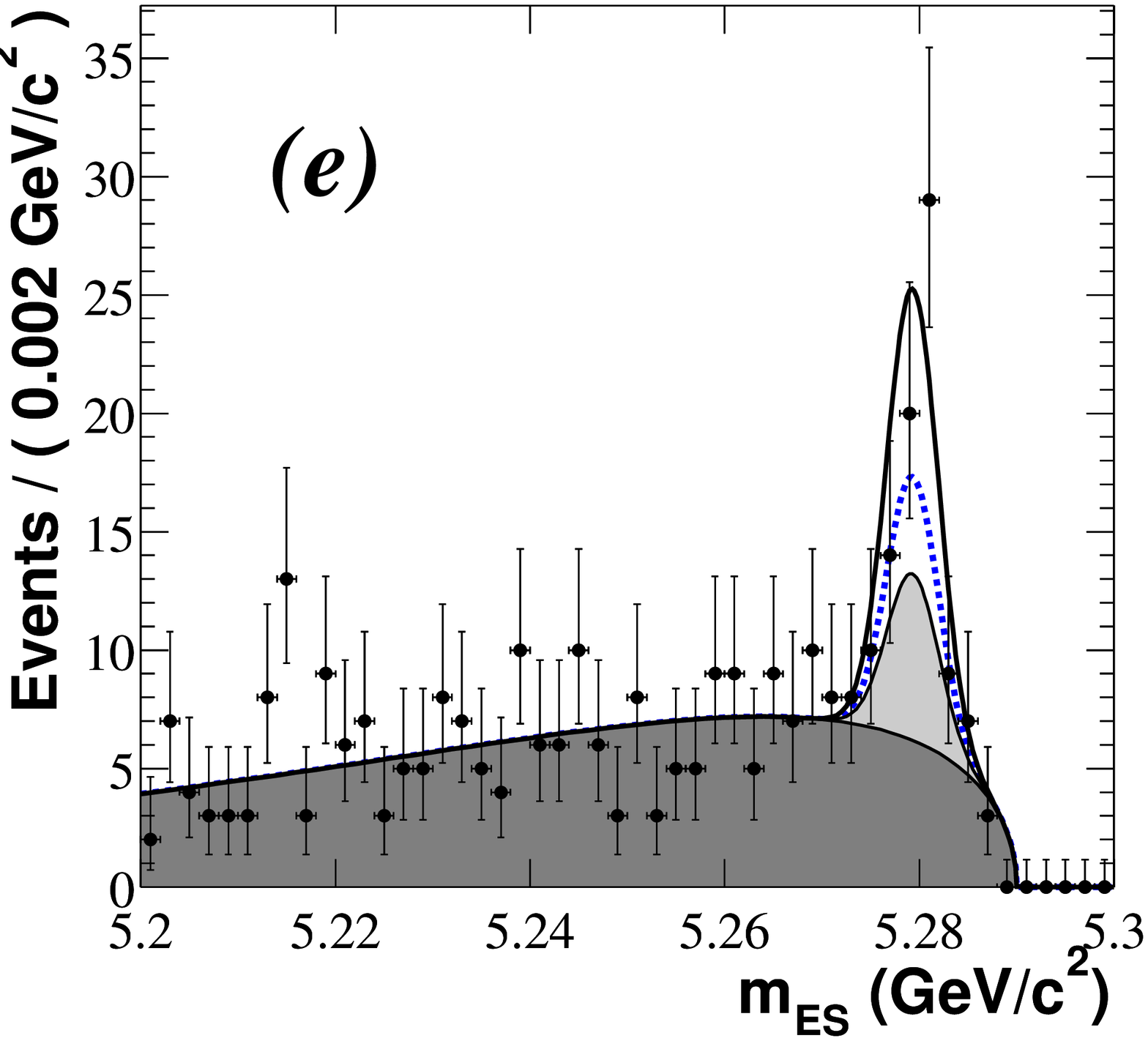,width=4.2cm}
\epsfig{file=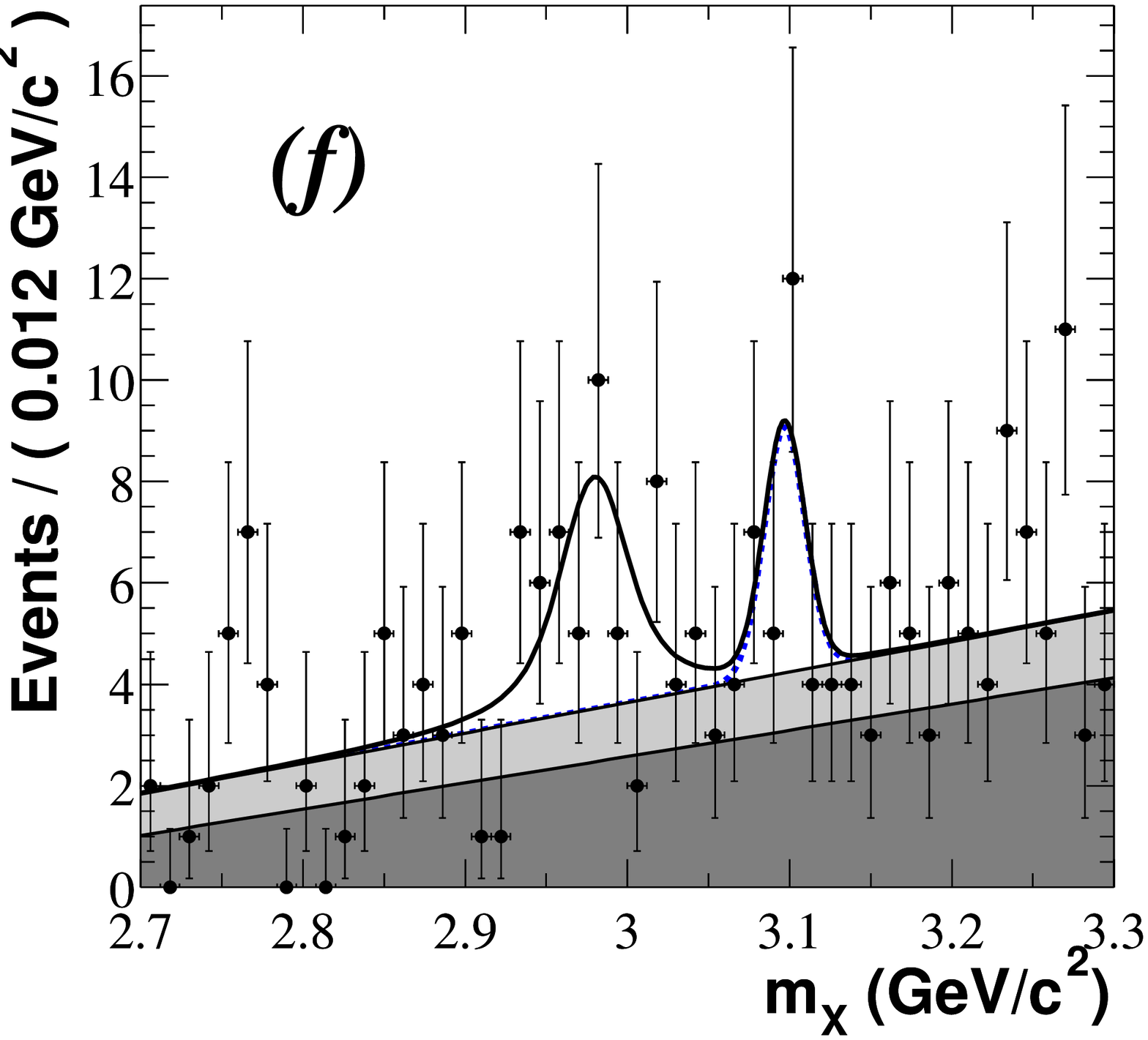,width=4.2cm}}
\mbox{\epsfig{file=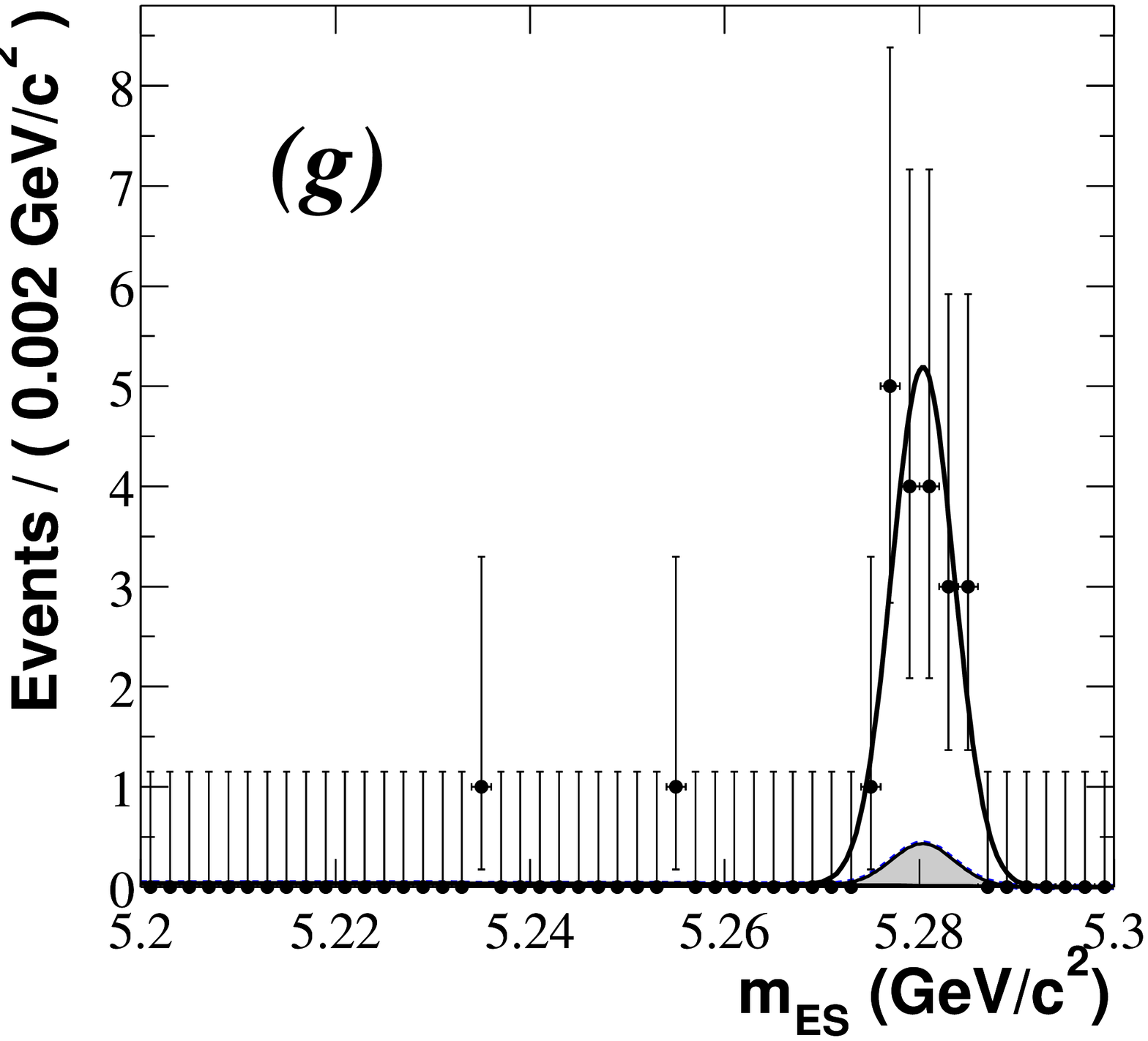,width=4.2cm}
\epsfig{file=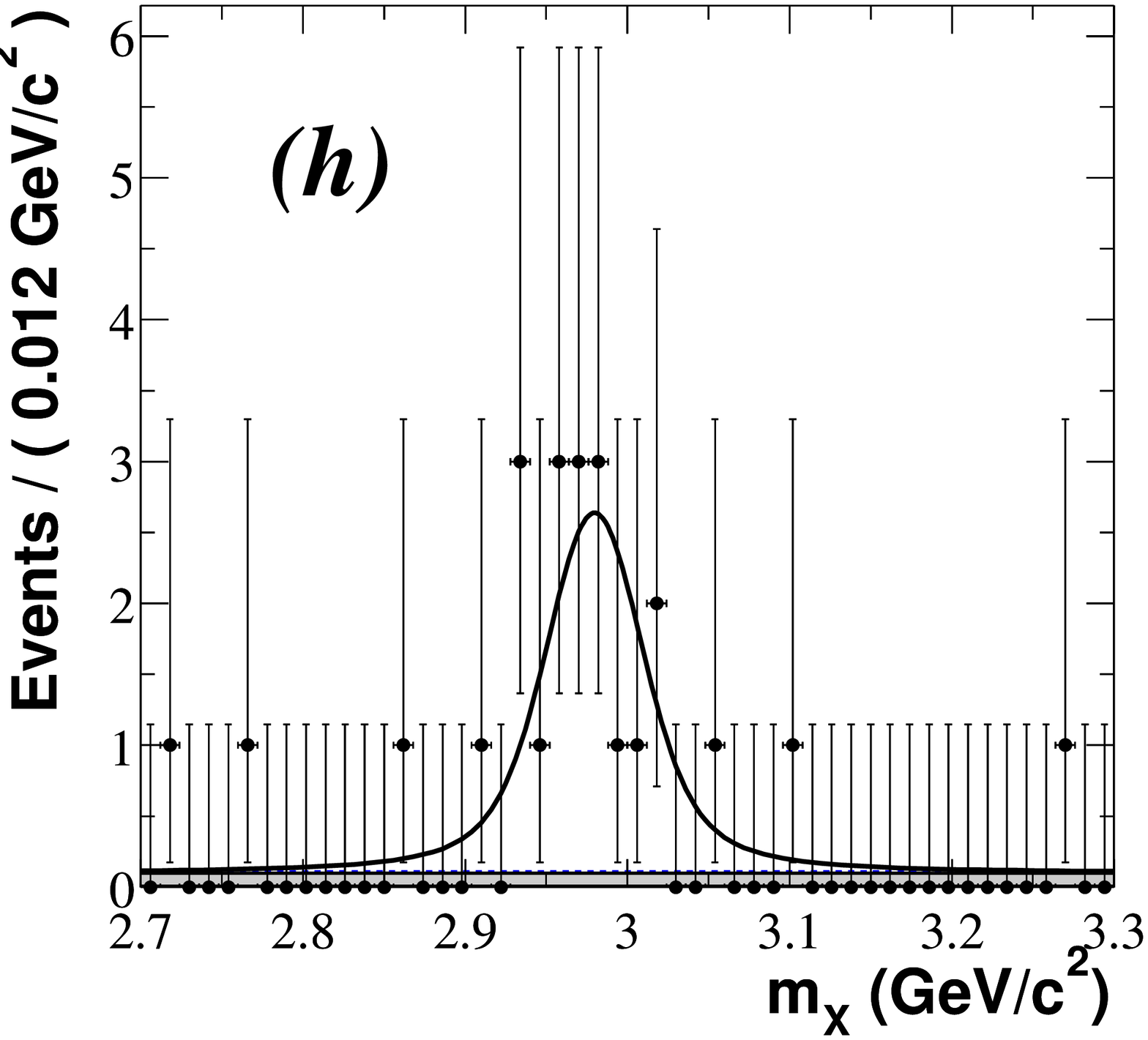,width=4.2cm}}
\end{center}
\caption
{Distributions of \mes (left) and \mX (right) for charged \B candidates. 
The \mes distributions displayed here are restricted to the 
2.90~$<\mX<$~3.15~\gevcc range; similarly, the \mX distributions include 
only events in the \mes signal region ($\mes>$~5.27~\gevcc). Each pair 
corresponds to a different \etac decay mode: (a, b) \eKsKPi; (c,d) \eKKPi;
(e,f) \shorteFourK; and (g,h) \eTwoPhi. The fitted p.d.f. projections are 
shown as solid curves. In each plot, the dark grey region corresponds to 
the combinatorial background component, light grey highlights the peaking 
background, and the dotted line is the sum of the total background and of 
the \jpsi component.}
\label{fig:allmodes}
\end{figure}

The measured \etac\ signal yields are reported in Table~\ref{tab:yields}. 
\begin{table}[!h]
\begin{center}
\caption{Number of \etacK events and statistical significance ${\cal S}$,
defined as
{\ensuremath{\sqrt{2 {\rm log} ({\cal L}_{\rm max}/{\cal L}_0)}}} where 
${\cal L}_{max}/{\cal L}_0$ is the likelihood ratio of the fitted maximum 
over the null hypothesis.
The first error is statistical, 
the second is the systematic uncertainty associated with the fitting 
procedure. 
}
\begin{tabular}{lcc} \hline \hline
Mode & \etac yield & ${\cal S}$ \\
\hline 
\etacKP &&\\
~~\eKsKPi & ~~306.4 \PM 24.4 \PM 14.0~~ &~~20.5 \\ 
~~\eKKPi  & ~~136.8 \PM 17.5 \PM 9.3~~  & ~~11.7 \\
~~\shorteFourK & ~~26.2  \PM 8.4 \PM 4.5~~ & ~~4.5\\
~~\eTwoPhi& ~~19.1  \PM 4.9 \PM 0.6~~ & ~~6.6 \\ \hline
\etacKN &&\\
~~\eKsKPi & ~~79.4 \PM 12.7\PM 4.3~~ & ~~9.7 \\
~~\eKKPi  & ~~40.9 \PM 9.5 \PM 2.7~~ & ~~6.2 \\
~~\shorteFourK &  ~~3.9 \PM 3.7 \PM 1.6~~ & ~~1.4 \\
~~\eTwoPhi&  ~~3.0 \PM 1.7 \PM 0.1~~ & ~~3.6 \\
\hline
\hline
\end{tabular}
\label{tab:yields}
\end{center}
\end{table}
We observe a significant signal in all modes with the exception of 
\Bz with \etac\ decaying into \shortFourK and \TwoPhi. 
The \mes and \mX 
distributions of \etacKP candidates are shown in Fig.~\ref{fig:allmodes}. 
In the largest samples (\eKsKPi) we can determine the \etac\ width 
$\Gamma(\etac )$ from a simultaneous fit to neutral and charged \B data. We 
find $\Gamma(\etac)$~=~39.7~\PM~6.6~\mevcc, where the error is statistical 
only, consistent with the \babar\ measurement, 
$\Gamma(\etac)$~=~34.3~\PM~2.3~\PM~0.9~\mevcc~\cite{gamma}.

The systematic uncertainty associated with the fitted signal yield includes 
three components: the uncertainty in the fixed parameters, the uncertainty 
associated with the Dalitz weighting procedure, and the uncertainty 
associated with the p.d.f. models. The first component is evaluated by 
varying each fixed parameter, one at a time, by one standard deviation and 
repeating the fit. This component is dominated by the uncertainty on 
$\Gamma(\etac)$ (0--3\% fractional uncertainty in \BR, depending on the mode).  
For the second component, the fit is repeated without applying the 
Dalitz-correction procedure; half of the difference on the \etac signal 
yield (0--2\% in \BR) is conservatively assigned as the corresponding systematic 
uncertainty. The last component is dominated by the uncertainty in the 
peaking background model. This error is evaluated by varying the assumed 
\mX dependence from a first- to a second-order polynomial; it typically 
amounts to 4\%, and exceeds 10\% only for the \shorteFourK\ modes. The 
error associated with the \mX resolution function model (0--5\%) is 
estimated by using, instead of a single Gaussian function fitted to the 
data, double-Gaussian resolution functions fitted to each simulated signal 
sample. 

Efficiencies are computed with simulated signal events that are 
reconstructed and selected using the same procedure as for the data, 
including the yield-extraction fit. 
\begin{table*}[!htbp]
\begin{center}
\caption{Efficiencies and relative systematic uncertainties.
}
\begin{tabular}{llcccclcccc} \hline \hline
     && \multicolumn{4}{c}{\etacKP} & \hspace{1cm}& \multicolumn{4}{c}{\etacKN} \\ 
     && \KsKPi & \KKPi & \shortFourK & \TwoPhi & \hspace{1cm} & \KsKPi & \KKPi & \shortFourK & \TwoPhi \\ \hline
&& \multicolumn{9}{c}{Signal efficiency} \\ 
&& 0.213 & 0.124 &  0.155 & 0.194 && 0.184 & 0.126 & 0.147 & 0.170 \\ \hline
Source of uncertainty && \multicolumn{9}{c}{Relative uncertainty on signal efficiency(\%)} \\ 
\cline{1-1} \cline{2-11}
Monte Carlo statistics      && 1.0  & 1.5 &  1.2 & 1.1 && 1.2 & 1.3 & 1.2 & 
1.1 \\
Tracking                    && 6.0  & 3.4 &  6.0 & 6.0 && 7.8 & 5.2 & 7.8 & 
7.8 \\
\KS reconstruction          && 3.0  &  -  &   -  &  -  && 6.0 & 3.0 & 3.0 & 
3.0 \\         
Particle identification     && 3.9  & 6.5 &  12.1& 8.0 && 1.5 & 3.9 & 9.4 & 
5.9 \\ 
\piz reconstruction         &&  -   & 5.0 &   -  &  -  &&  -  & 5.0 &  -  &  
-  \\  
Selection cuts              && 2.7  & 2.8 &  1.7 & 2.2 && 3.3 & 3.5 & 3.1 & 
3.4 \\ 
Yield-extraction fit        && 3.0  & 3.0 &  3.0 & 3.0 && 3.0 & 3.0 & 3.0 & 
3.0 \\ 
\hline
Total uncertainty           && 8.8  & 9.8 & 13.9 & 10.7&& 10.9& 9.9 & 13.3& 
11.2\\   
\hline
\hline

\end{tabular}
\label{tab:efficiency}
\end{center}
\end{table*}
We apply small corrections, determined from data, to the efficiency 
calculation to account for the overestimation of the tracking and 
particle-identification performance, and of the \piz and \KS reconstruction 
efficiencies. A systematic uncertainty is assigned to each correction to 
account for the limited size and purity of the control sample used in 
computing that correction. For example, for the fast kaon identification, 
we correct the simulation using a pure sample of 
$\Dstarp\rightarrow\pip\Dz$ decays with $\Dz\rightarrow\Km\pip$. We include 
in the particle-identification systematic uncertainty contributions 
associated with the sample size, the background subtraction, and the 
different kinematics of this decay chain compared to the two-body \etacKP 
decay. Similarly, corrections affecting the \piz reconstruction are 
calibrated using real and simulated $\epem \ra \tau^+ \tau^-$ and 
multihadron samples.

In addition, after all corrections, we compare our signal simulation to 
appropriate control samples with similar kinematics or final-state 
topology, in order to quantify the ability of the simulation to model the 
kinematic and event-shape variables used in the event selection. The small 
residual differences in the efficiencies at the cut value are assigned as 
systematic uncertainties affecting the selection procedure.

Finally, we assign a systematic uncertainty to the yield-extraction 
fit by 
evaluating the influence of mixing background events with simulated signal 
events. Values for the efficiencies, the corrections, and the corresponding 
systematic uncertainties are reported in Table~\ref{tab:efficiency}.

The results on the products of the branching fractions for each mode are 
listed in Table~\ref{tab:products}.
\begin{table}[htbp]
\caption{Measured branching-fraction products
\BR(\etacK)$\times$\BR(\etac$\rightarrow X$) (10$^{-6}$). The first error is 
statistical and the second is the total systematic uncertainty.}
\begin{center}
\begin{tabular}{lrr} \hline \hline
\etac decay channel & \etacKP   & \etacKn  \\ \hline 
\eKzKPi          & 48.6 \PM 3.9 \PM 4.9 &~~42.6 \PM 6.8 \PM 5.2  \\
\eKKPi           & 12.9 \PM 1.7 \PM 1.6 & 11.1 \PM 2.6 \PM 1.3 \\
\shorteFourK          & 2.0  \PM 0.6 \PM 0.4  & 0.9  \PM 0.9 \PM 0.4 \\
\eTwoPhi         & 4.7  \PM 1.2 \PM 0.5  & 2.4  \PM 1.4 \PM 0.3 \\
\hline \hline
\end{tabular}
\end{center}
\label{tab:products}
\end{table}
We use the world-average values for the \ksdk, \pizdk and \phidk branching 
fractions~\cite{PDG} and include their uncertainties in the systematic 
error. The systematic error also comprises the uncertainties from the 
determination of the number of \BB pairs (1.1\%), from the likelihood fit, 
and from the signal efficiency. We assume that the branching fraction of 
the \FourS into \BB is 100\%, with an equal admixture of charged and 
neutral \B final states. We do not include any additional uncertainty due 
to these assumptions. Possible interference effects 
between the \etacK signal and the peaking background are neglected.

The decay amplitudes for \eKKPi and \eKzKPi are related by isospin 
symmetry. The expected ratio of branching fractions, using the appropriate 
Clebsch-Gordon coefficients, is 0.25. Our measurements are consistent with 
this value for both the \Bp (0.27 \PM 0.04 \PM 0.03) and the \Bz (0.26 \PM 
0.07 \PM 0.03) sample. We therefore combine our results for these modes and 
use the world average for the \eKKbarPi branching fraction (0.055 \PM 0.017~\cite{PDG}) to 
derive 
$$\BR(\etacKP) = (1.34 \pm 0.09 \pm 0.13 \pm 0.41) \times 10^{-3} $$
$$\BR(\etacKn) = (1.18 \pm 0.16 \pm 0.13 \pm 0.37) \times 10^{-3} $$
where the first error is statistical, the second systematic, and the third 
due to the uncertainty on the \eKKbarPi branching fraction. In the 
combination we separate correlated and uncorrelated uncertainties to weight 
the individual results and obtain the total systematic error. We also 
compute the ratio of neutral over charged \B decays 
$ \BR(\etacKn)/\BR(\etacKP) = 0.87 \pm 0.13 \pm 0.07 $,
and, multiplying by the mean lifetime ratio $\tau_{B^+}/\tau_{B^0} = 1.085 
\pm 0.017 $~\cite{PDG}, we derive the ratio of partial widths
$$ \Gamma(\etacKn)/\Gamma(\etacKP) = 0.94 \pm 0.14 \pm 0.08.$$

To determine $R_K = \Gamma(\etacK)/\Gamma(\jpsiK)$, we use the \babar\ 
measurements~\cite{ExclBabar} of the branching fractions, 
\BR($\Bp \rightarrow \jpsi \Kp$)~=~$(10.1 \pm 0.3 \pm 0.5)\times 10^{-4}$ 
and 
\BR($B^0 \rightarrow \jpsi K^0$)~=~$(8.5 \pm 0.5 \pm 0.6)\times 10^{-4}$.
We obtain
$$R_K (B^+) = 1.33 \pm 0.10 \pm 0.12 \pm 0.41$$
$$R_K (B^0) = 1.39 \pm 0.20 \pm 0.13 \pm 0.43,$$
where the first error is statistical, the second systematic, and the third 
due to \eKKbarPi branching fraction. Our results agree with most 
predictions for $R_K$, which range from 0.9 to 
2.3~\cite{QCD,RK1,RK2,RK3,RK4}.

The measured values of \BR(\shorteFourK) and \BR(\eTwoPhi) have higher 
uncertainties and therefore these modes are not used for averages. We can 
express our \shorteFourK and \eTwoPhi results in terms of ratios to the 
best-measured branching fractions of \eKKbarPi, thereby cancelling all 
fully-correlated systematic uncertainties. We average results on charged \B 
decays and neutral \B decays, taking into account correlations in the 
systematic uncertainties, to obtain
$\BR(\shorteFourK)/\BR(\eKKbarPi) = (2.3 \pm 0.7 \pm 0.6) \times 10^{-2}$ 
and
$\BR(\eTwoPhi)/\BR(\eKKbarPi) = (5.5 \pm 1.4 \pm 0.5) \times 10^{-2}$.
These results can be translated into \etac\ branching fractions:
$$ \BR(\shorteFourK) = (1.3 \pm 0.4 \pm 0.3 \pm 0.4)\times 10^{-3}$$
$$ \BR(\eTwoPhi)= (3.0 \pm 0.8 \pm 0.3 \pm 0.9)\times 10^{-3},$$
where the third error is due to the uncertainty of \BR(\eKKbarPi). Note 
that about half of the \shorteFourK events are due to
\eTwoPhi, \phidk decays.
Our measured branching fractions for \shorteFourK and \eTwoPhi are 
consistent with recent results from Belle and BES~\cite{4kBelle,2phiBES}
and are smaller than those of earlier experiments~\cite{PDG}.

As a cross-check, we can extract the branching fraction of \jpsi decaying 
into the \shortFourK final state from the measured number of \jpsi events 
in the appropriate \Bp and \Bz samples. 
Assuming the same efficiencies as 
for the \etacK (\shorteFourK) processes and using the \babar\ measurements 
of \BR($B^+ \rightarrow \jpsi K^+$) and of 
\BR($B^0 \rightarrow \jpsi K^0$) \cite{ExclBabar}, we obtain 
$ \BR(J/\psi \rightarrow\shortFourK)= (1.0 \pm 0.5)\times~10^{-3}~~(\Bp)$ 
and
$ \BR(J/\psi \rightarrow\shortFourK)= (0.1 \pm 0.2)\times~10^{-3}~~(\Bz)$, 
where the error is statistical only. These results are consistent with the 
world average, 
$\BR(J/\psi \rightarrow\shortFourK)$= (0.7 $\pm$ 0.3 ) $\times$ 10$^{-
3}$~\cite{PDG}.
 
In summary, we have studied \etacK decays with \etac decaying into \KsKPi, 
\KKPi, \shortFourK, and \TwoPhi. Using the first two decay channels, we 
have measured the branching fractions 
$\BR(\etacKP) = (1.34 \pm 0.09 \pm 0.13 \pm 0.41) \times 10^{-3} $ and
$\BR(\etacKn) = (1.18 \pm 0.16 \pm 0.13 \pm 0.37) \times 10^{-3} $,
which improve the statistical precision of, and are in good agreement with, 
previous measurements~\cite{EtacCLEO,EtacBelle}. We have also measured the 
branching-fraction ratios
 $\BR(\shorteFourK)/\BR(\eKKbarPi) = (2.3 \pm 0.7 \pm 0.6) \times 10^{-2}$ 
and
$\BR(\eTwoPhi)/\BR(\eKKbarPi) = (5.5 \pm 1.4 \pm 0.5) \times 10^{-2}$,
where \shorteFourK includes \eTwoPhi events with \phidk. 
The inferred  branching fractions of \shorteFourK and \eTwoPhi are
in good agreement with recent results and smaller than suggested
by earlier experiments.

We are grateful for the excellent luminosity and machine conditions
provided by our \pep2\ colleagues, 
and for the substantial dedicated effort from
the computing organizations that support \babar.
The collaborating institutions wish to thank 
SLAC for its support and kind hospitality. 
This work is supported by
DOE
and NSF (USA),
NSERC (Canada),
IHEP (China),
CEA and
CNRS-IN2P3
(France),
BMBF and DFG
(Germany),
INFN (Italy),
FOM (The Netherlands),
NFR (Norway),
MIST (Russia), and
PPARC (United Kingdom). 
Individuals have received support from the 
A.~P.~Sloan Foundation, 
Research Corporation,
and Alexander von Humboldt Foundation.


\begin{thebibliography}{99}

\bibitem{sin2bBabar} \babar\ Collaboration, B. Aubert {\it et al.}, \jprl{89}, 201802 (2002).
\bibitem{sin2bBelle} Belle Collaboration, K. Abe {\it et al.}, \jprd{66}, 071102 (2002).
\bibitem{QCD} N. G. Deshpande and J. Trampetic, \plb{339}, 270 (1994).
\bibitem{RK1} M. R. Ahmady and R. R. Mendel, \zpc{65}, 263 (1995). 
\bibitem{RK2} P. Colangelo, C. A. Domingues, and N. Paver, \plb{352}, 134 (1995).
\bibitem{RK3} M. Gourdin, Y. Y. Keum, and X.-Y. Pham, \jprd{52}, 1597 (1995).
\bibitem{RK4} D. S. Hwang and G.-H. Kim, \zpc{76}, 107 (1997).
\bibitem{Conjunote} Charge conjugate modes are implicity included throughout this paper.
\bibitem{NIM} \babar\ Collaboration, B. Aubert {\it et al.}, \nima{479}, 1 (2002).
\bibitem{PDG} Particle Data Group, K. Hagiwara {\it et al.}, \jprd{66}, 010001 (2002)
   and 2003 off-year partial update for the 2004 edition available on the PDG WWW pages
   (URL: http://pdg.lbl.gov/).
\bibitem{FisherCLEO} CLEO Collaboration, D. M. Asner {\it et al.}, \jprd{53}, 1039 (1996).
\bibitem{ARGUS} ARGUS Collaboration, H. Albrecht {\it et al.}, \zpc{48}, 543 (1990).
\bibitem{gamma} \babar\ Collaboration,  B. Aubert {\it et al.}, hep-ex/0311038, submitted to \jprl.
\bibitem{ExclBabar} \babar\ Collaboration, B. Aubert {\it et al.}, \jprd{65}, 032001 (2002).
\bibitem{4kBelle} Belle Collaboration, H.-C. Huang {\it et al.}, \jprl{91}, 241802 (2003).
\bibitem{2phiBES} BES Collaboration, J. Z. Bai {\it et al.}, \plb{578}, 16 (2004).
\bibitem{EtacCLEO} CLEO Collaboration, K. W. Edwards {\it et al.}, \jprl{86}, 30 (2001).
\bibitem{EtacBelle} Belle Collaboration, F. Fang {\it et al.}, \jprl{90}, 071801 (2003).
\end{thebibliography}
\end{document}